\documentclass[manuscript]{acmart} 

\usepackage{subcaption}
\usepackage{caption}
\usepackage{wrapfig}
\usepackage{multirow}
\usepackage[para]{footmisc}

\newcommand{\ourmethod}{\textit{Stungage}}

\AtBeginDocument{%
  \providecommand\BibTeX{{%
    \normalfont B\kern-0.5em{\scshape i\kern-0.25em b}\kern-0.8em\TeX}}}

\copyrightyear{2022}
\acmYear{2022}
\setcopyright{acmcopyright}\acmConference[UMAP '22]{Proceedings of the 30th ACM Conference on User Modeling, Adaptation and Personalization}{July 4--7, 2022}{Barcelona, Spain}
\acmBooktitle{Proceedings of the 30th ACM Conference on User Modeling, Adaptation and Personalization (UMAP '22), July 4--7, 2022, Barcelona, Spain}
\acmPrice{15.00}
\acmDOI{10.1145/3503252.3531307}
\acmISBN{978-1-4503-9207-5/22/07}

\begin{document}

\title[\textit{I Cannot See Students Focusing on My Presentation ...}]{\textit{I Cannot See Students Focusing on My Presentation; Are They Following Me?} Continuous Monitoring of Student Engagement through ``\textit{Stungage}''}

\author{Snigdha Das, Sandip Chakraborty, Bivas Mitra}
\affiliation{%
	\institution{Department. of Computer Science and Engineering, Indian Institute of Technology Kharagpur, India}
}
\email{snigdhadas@sit.iitkgp.ac.in, {sandipc,bivas}@cse.iitkgp.ac.in}

\begin{abstract}
Monitoring students' engagement and understanding their learning pace in a virtual classroom becomes challenging in the absence of direct eye contact between the students and the instructor. Continuous monitoring of eye gaze and gaze gestures may produce inaccurate outcomes when the students are allowed to do productive multitasking, such as taking notes or browsing relevant content. This paper proposes {\ourmethod} -- a software wrapper over existing online meeting platforms to monitor students' engagement in real-time by utilizing the facial video feeds from the students and the instructor coupled with a local on-device analysis of the presentation content. The crux of {\ourmethod} is to identify a few opportunistic moments when the students should visually focus on the presentation content if they can follow the lecture. We investigate these instances and analyze the students' visual, contextual, and cognitive presence to assess their engagement during the virtual classroom while not directly sharing the video captures of the participants and their screens over the web. Our system achieves an overall F2-score of $0.88$ for detecting student engagement. Besides, we obtain $92$ responses from the usability study with an average SU score of $74.18$. 
\end{abstract}


\maketitle

\section{Introduction}
The pandemic has made virtual online classes a norm rather than an exception. However, the online mode of classes has received several criticisms; one of the major criticisms being it lacks the eye-contact between the teacher (or the instructor) and the students. In a classroom, such eye contacts significantly help the instructor gauge the students' learning pace and understand whether the students are engaged with the topic being taught. In the era of pandemic, a large number of studies~\cite{bace2020far,whitehill2014faces,kamath2016crowdsourced,robal2018webcam,aslan2019investigating,mohamad2019automatic,jensen2020toward} have highlighted this requirement. Consequently, several works have utilized signals like video captured from the front camera~\cite{kar2020gestatten,bace2020quantification,he2021gazechat} or utilized specialized devices like smart glasses, thermal cameras, eye-trackers, etc.~\cite{whitehill2014faces,vspakov2019two,abdelrahman2019classifying,xiao2017undertanding,herbig2020investigating} to capture the eye dynamics of the students to analyze how they interact with the computer during a live lecture. Intuitively, a solution involving such specialized devices can not scale well for the masses, whereas a continuous eye monitoring-based solution poses a major limitation as follows. 

Interestingly, a virtual classroom opens up the scope for multitasking~\cite{marlow2016taking,cao2021large,chen2021learning,benke2021leadboski}, where a student may perform several other activities while still attending the classes online. These activities range from productive activities that support interaction with the classroom during the live lecture (like taking notes, browsing related concepts on the web, etc.) to the activities that negatively impact the attentiveness towards the classroom (like browsing social media pages, chatting over the phone, etc.). In both the above cases, the eyes of a student might not be focused on the computer's screen; a method that solely analyzes eye dynamics to infer students' engagement may result in false positives when the student performs productive multitasking. Understanding students' attention in the presence of multitasking is challenging, as a student might get involved with such activities for a significant duration during a live class~\cite{cao2021large}. Apart from that, gazing at the screen is not an essential condition for getting involved in an online meeting~\cite{he2021you,george2022users}. As shown in several recent studies, a student might still get actively involved in a virtual classroom even if they minimally gaze at the screen~\cite{gao2021digital,barmaki2015providing,seo2019joint,das2021quantifying}. Therefore, we argue that continuous tracking of eye gazes does not provide a reliable source of information for marking a student inattentive in a virtual classroom. 

Consequently, we ask the following question in this paper: \textit{how can we quantify a participants' engagement while allowing free movements and other activities that promote positive multitasking?} Finding a generic solution for this problem is challenging, and the pedagogy changes depending on multiple factors, like the level of teaching (K-12 or University), subjects and topic, socio-cultural aspects, etc. This paper focuses on a particular case when the teacher utilizes a presentation or slides to explain the concept. The core idea is that a presentation with textual and animated slides often triggers intermediate cues when the meeting participants are tempted to look at the screen if they are attentive. We call these cues the \textit{Fixation Target Events} which include a figure or a diagram, animations, highlighted texts, etc. Even with this specific setup, multiple technical challenges need to be addressed. \textbf{First}, the processing needs to be in real-time on the video feed over the meeting platform. \textbf{Second}, it might happen that the student is browsing his social media profile during the virtual classroom. In this case, his eye gaze on the screen will also be captured during the fixation target events, resulting in spurious false positives. The platform needs to analyze whether the gaze is on the presentation slide or on his social media profile. \textbf{Third}, a naive approach of understanding whether a student focuses on the same content that the instructor is presenting would be to compare the screen of the student with that of the instructor. However, processing the video frames from the instructor as well as all the students and comparing them in real-time is challenging. Further, the instructor and the student might use different devices having different screen sizes; therefore, a direct comparison might be difficult. It can also be noted that a meeting application should record the minimum information about the participant's screen such that the privacy of the participants is preserved.

\subsection{Our Contributions}
\begin{wrapfigure}{l}{0.5\linewidth}
	\centering
	\includegraphics[clip,width=\linewidth,keepaspectratio]{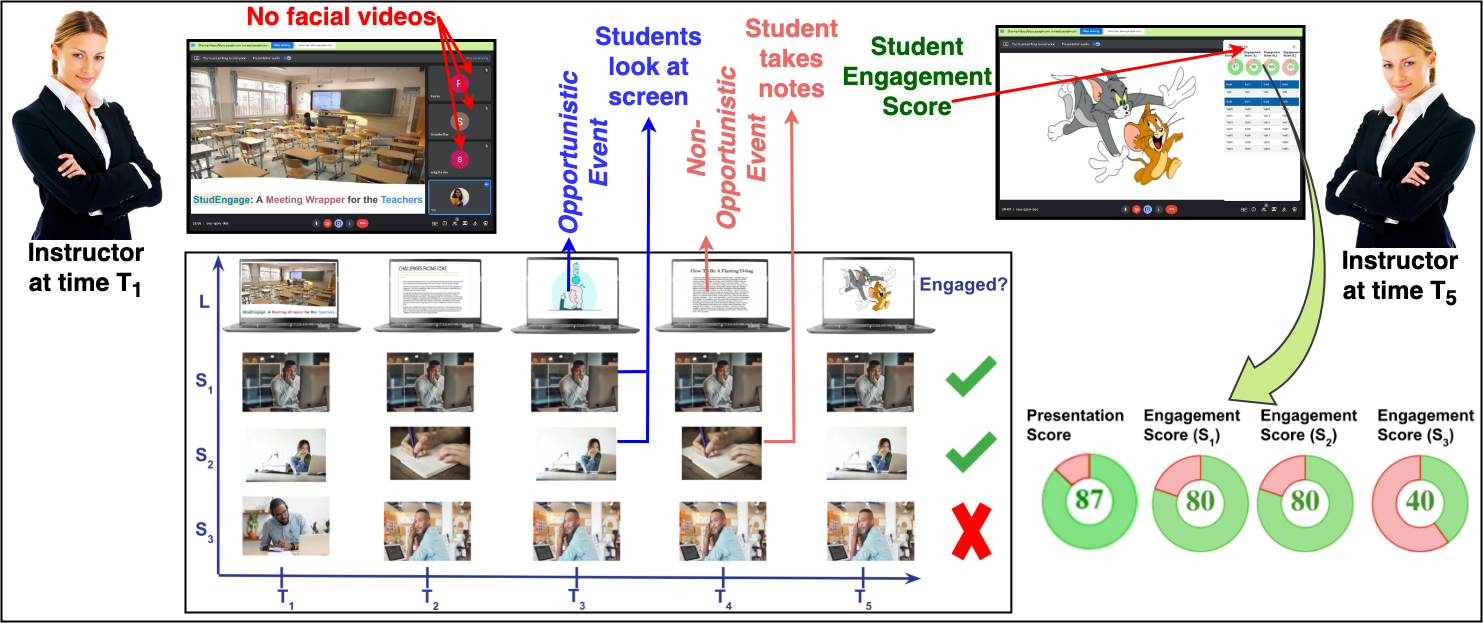}
	\caption{We propose {\ourmethod} to detect students' engagement in the virtual classroom using the pervasive webcam. Our method locally analyses both the students and the instructor's video feed along with the lecture presentation and finally compares at instructor's end to infer the engagement.}
	\label{fig:engageSystem}
\end{wrapfigure}

Owing to these challenges and limitations of the prior works, we propose {\ourmethod} -- a student engagement detection system that aims to capture both the students and the instructor's video feed along with the lecture presentation to infer the involvement of the students in the virtual classroom (Figure \ref{fig:engageSystem}). {\ourmethod} works as a software wrapper on top of an online meeting platform where both the instructor's and the students' video feeds are processed locally. The computed information is shared with the instructor for generating an involvement score for each of the students. The core contributions of this paper are as follows.

\noindent\textbf{(1) Detection of fixation target events:}
The fundamental premise of our work is that even if a student involves in multitasking, the attentive one fixates on the fixation target points such as animation, image, and highlighted short text content. Accordingly, {\ourmethod} extracts the target points from the lecture video by detecting the foreground object movement followed by a Spatio-temporal bound measure. 

\noindent\textbf{(2) Analyzing student's understandability:}
For understanding the students learning pace, {\ourmethod} uses a cascade-like phenomenon while responding to three questions -- (1) \textit{are you inside the online class?}, which detects the visual presence of the students during the fixation target events, (2) \textit{are you looking at the presentation?}, which detects the contextual presence of the students by mapping the presence of the instructor and student, and (3) \textit{are you following the presentation? }, which finally detects the cognitive presence of the students by comparing the instructor and the student's gazing energy at the screen. We capture the visual presence by \textit{frontal face detection mechanism} to segregate the activities like watching mobile, browsing Facebook, sleeping, etc., from following the lecture presentation by developing a novel method of extracting the spectral properties of the gazing histogram. 

\noindent\textbf{(3) Analyzing teaching performance:}
{\ourmethod} computes the instructor's presentation score as a by-product of the system. We count the instructor's visual presence during the fixation target event, and finally, upon aggregation over a time window, the \textit{presentation score} is generated.

\noindent\textbf{(4) Prototype deployment \& evaluation:}
We have developed a prototype of {\ourmethod} and tested it over two different studies -- (i) a pilot study both in lab and in-the-wild set up to investigate the system performance over the existing systems, (ii) a usability study to test the usability of the system. We have recruited $30$ participants belonging to the age group of $24$-$44$ years to perform both the pilot experiments. We achieve an overall F2-score of $0.88$ for detecting student engagement. On contrary, we obtain $92$ responses from the usability study with an average SU score of $74.18$.

\section{Related Work}
Existing literature primarily focuses on three different strategies for detecting student engagement in a classroom -- (1) questioning-based, (2) dedicated device-based, and (3) commonly off-the-shelf device-based approach. 

\noindent\textbf{Questioning-based approach:} 
Similar to the physical classroom system, for understanding the students' learning space, the question-answering interaction-based solution~\cite{shin2018understanding,winkler2020sara,price2020engaging,yeckehzaare2020qmaps,morshed2021progression} is one of the traditional ways in the virtual classroom system. In \cite{shin2018understanding}, Shin \emph{et al.} studied the instructor and the learner perceptions using the in-video prompting questionnaire. Besides, Price \emph{et al.} \cite{price2020engaging} applied a comparison mechanism for detecting the engagement of the students where the instructor's solution was provided, and they were prompted to compare their solution with the instructor's one. In separate work, Yeckehzaare \emph{et al.} \cite{yeckehzaare2020qmaps} used the concept of question generation and linking by applying a question map for engaging the students. In these cases, the students proactively participate in the different forms of questionnaires to establish their understanding. Apart from the questionnaire, the voice and text-based interaction \cite{winkler2020sara} also plays a significant role in improving learning in online education.

\noindent\textbf{Dedicated device-based approach:} 
To address the problem of the student involvement in the virtual classroom, several studies~\cite{sharma2015displaying,d2016gazed,avrahami2016supporting,xiao2017undertanding,d2018eye,yao2018visualizing,kosch2018your,pham2018adaptive,srivastava2019continuous,kutt2019eye,abdelrahman2019classifying,vspakov2019two,landsmann2019classification,babaei2020faces,kutt2020effects,sims2020neural,ueno2020estimating,subburaj2020multimodal,herbig2020investigating} have explored the use of dedicated devices for capturing either the behavioral or the physiological signals of the students. In one of the earliest studies, Sharma \emph{et. al.} \cite{sharma2015displaying} tried to capture the students' lecture navigation pattern by displaying the instructor's gaze. The researcher observed that showing the gaze made the presentation easier for the students following the lecture. Afterwards, a few works \cite{d2016gazed,d2018eye,kutt2019eye,vspakov2019two} captured the eye gaze signal through the eye tracker for collaborative reading, writing, problem-solving, and learning. Later on, the authors in~\cite{kutt2020effects,subburaj2020multimodal} explored acoustic signal along with the eye gaze for improving the remote collaborative performance. While eye gaze monitoring by eye tracker is a promising technique, the cost and availability of the tracker to all the students is a major obstacle. To address these issues, different forms of other sensing such as thermal imaging~\cite{srivastava2019continuous,abdelrahman2019classifying}, mouse \& keyboard tracking~\cite{avrahami2016supporting}, and PPG~\cite{xiao2017undertanding} are used to capture the physiological signal to infer the attentiveness during the lecture session. 
Despite the benefits of physiological sensing, it is commonly observed that the techniques require the continuous intervention of the attendees, which is typically not feasible during the lecture session as students can forget to track the signal. Furthermore, the dedicated wearable devices need special attention towards installing and demonstrating the devices, which is not a preferable resolution for a large class.

\noindent\textbf{Commonly off-the-shelf device-based approach:} 
To suppress the shortcomings of the dedicated invasive devices used in the virtual classroom, the pervasive webcams are considered a suitable alternative for capturing the attendees' gaze signature. For instance, Whitehill \emph{et al.} \cite{whitehill2014faces} studied the student engagement in the context of their facial expression. In the same line, authors in~\cite{soltani2018facial,aslan2019investigating,sun2019presenters,mohamad2019automatic} applied various emotional attributes such as satisfied, confused, bored, and anxiety for detecting the involvement of the students in the virtual classroom. While keeping the emotion detection in the context of engagement is a promising way; however, the frontal screen with the lecture content is one of the mandate criteria for processing the data. The attendee can look at different content and give similar expressions. Additionally, in the absence of the instructor's expression, the attendees can give different expressions irrespective of engaged or non-engaged. To address these limitations, some studies explores the gaze-based visual attention~\cite{bace2020far,kuzminykh2020classification,bace2020quantification,kar2020gestatten} for finding the attentiveness of the attendees. In~\cite{bace2020quantification}, Bace \emph{et al.} quantified the visual attention by checking whether the attendee was looking at the frontal screen. In~\cite{bace2019accurate,bace2020combining}, the authors further extended the work by comparing the screen object with the gaze projection on the screen. The research detected the pursuit interaction but also acknowledged that the objects on the screen were known and the screen was large. Kar \emph{et al.}~\cite{kar2020gestatten} compared the attendee's gaze gesture with the instructor's one to conclude the participants' attentiveness. However, all of these works consider continuous monitoring of students' gaze, which is impractical in multitasking and thus can yield severe false positives. 

Similar to the state-of-the-art, our work also uses pervasive webcams to monitor visual, contextual, and cognitive attention but explores all the attentional behavior simultaneously along with the consideration of discrete monitoring. Our research goal is to identify the different characteristics of the various attentional behavior and develop a system that shows the students' engagement in a real-time virtual classroom while allowing the student to multitask.

\section{The Design of {\ourmethod}: Core Idea and Broad System Overview}
{\ourmethod} infers the student involvement and teaching performance from the Spatio-temporal analysis of the student and the instructor video feeds and the lecture content. The system runs on the students' device that captures and processes the students' video feed at their end to produce the meta-data. The meta-data are compared at the instructor's device to detect the involvement of the students in the online lecture session. We start with a preparatory study that helps us understand the requirements for developing such as system. 

\subsection{Preparatory Study}
For establishing the requirement of detection of student engagement during online teaching, we have conducted an online anonymous survey\footnote{\url{https://lnkd.in/e3T9F\_d}} over $466$ teachers and students from different locations across the globe. The participants are from different designations, including undergrad students ($51.5\%$), masters students ($12.7\%$), research scholar ($18.8\%$), faculty ($10.3\%$), and so on. We found that most of our studied population ($87.4\%$) admit that online classes lose the charm of the physical classroom. Presentation slide-based teaching is one of the popular teaching modes in the virtual classroom, where animation, image, and highlighted text are the preferable presentation content. Thereby, the attentive students fixate on those contents. Asking about the video-sharing reveals that the participants ($65.6\%$) are comfortable sharing their videos when the audience size is small (within $20$). This motivates us to work with video extracted information sharing. We observe that $65.3\%$ of the participants strongly believe that multitasking is a common tendency during the online session, which introduces the need for the discrete interval local processing of video extracted information-sharing schemes\footnote{Due to the interest of space, we exclude the complete human study. However, the readers can check the details of this study through this link -- \url{https://github.com/Stungage/PreparatoryStudy}.}. This discrete computation involves the opportunistic events where the student fixates on the presentation. We process these ideas to develop our student engagement detection system.   

\subsection{Design Idea}
\begin{wrapfigure}{l}{0.6\linewidth}
	\centering
	\includegraphics[clip,width=\linewidth,keepaspectratio]{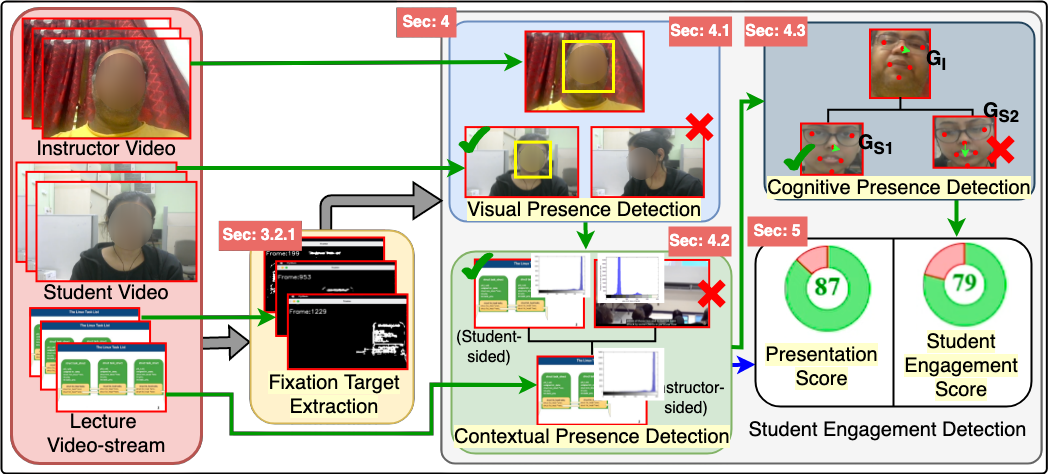}
	\caption{Student engagement detection framework modules --  \textit{Fixation Target Extraction} and \textit{Student Engagement Detection}. $G_I, G_{S1}, G_{S2}$: gazing energy of instructor, student 1 \& student 2, respectively.}
	\label{fig:engageSystemFramework}
\end{wrapfigure}

The overall idea of the system is to identify the opportunistic events where the attentive student must fixate on the screen and analyze the lecture context and gaze movement during those opportunistic events. We call these events the fixation target events. 
Figure \ref{fig:engageSystemFramework} shows the overall framework of the student engagement detection system, which is primarily composed of two modules -- (a) \textit{Fixation Target Extraction}, and (b) \textit{Student Engagement Detection}. The first module analyses the presentation video content from the lecture presentation to extract the fixation targets. The final module studies the presenter and the students' video feed during the fixation target events to detect the student engagement during the online presentation-based teaching. Additionally, the final outcome includes the presenter score as a by-product of the system for characterizing the instructor's performance in the session.

\subsubsection{Fixation Target Extraction} 
This module runs at both the instructor and the students' end and excerpts the fixation target points from the presentation video. This involves two steps. 

\noindent\textbf{(a) Foreground Video Extraction:} {\ourmethod} first identifies the object movements within the presentation slide. Without loss of generality, we assume that each presentation is made of a single template. Thus, the template represents the background of the entire presentation video feed and the variable content on top of the template appears as the foreground of the presentation video. Therefore, for filtering out the invariant component from the presentation video, {\ourmethod} applies the existing Gaussian mixture model-based background subtraction mechanism \cite{zivkovic2004improved}. However, due to the imprecise learning of the Gaussian parameters, the extracted foreground pixels incorporate scattered spots. {\ourmethod} relies on the median filtering on the foreground video for erasing the salt pepper-like scatter spots.

\noindent\textbf{(b) Fixation Target Detection:} This module considers the filtered foreground pixels to precisely detect the opportunistic events, called fixation target events. The key idea is to identify the portions of the lecture content where the attentive student fixates. Among the different presentation lecture video content, our study shows that animation, image, and short highlighted text give additional attention to the audience. Furthermore, along with the specified presentation content, the pointer movement creates attention towards the audience. {\ourmethod} detects these events by applying a Spatio-temporal threshold mechanism. The spatial threshold is applied to eliminate the text-heavy presentation content whereas the temporal threshold removes the short non-resistant presentation content. Specifically, an event is marked as a fixation target event when the foreground frame pixel count is within the spacial thresholds $\delta_{s_1}$ and $\delta_{s_2}$ and that spacial constraint persists at least for $\delta_{t}$ number of frames where $\delta_{t}$ is the temporal threshold. Any violation of the Spatio-temporal threshold marks the foreground selection as the non-fixation target event.

\subsubsection{Student Engagement Detection}
Partially, this module executes on both the instructor and the students' sides, and the rest runs on the instructor side to generate the student engagement scores during the fixation target events. This module works in a cascade-like phenomenon while responding to three questions -- (1) \textit{Are you inside the online class?}, detecting the visual presence of the students during the fixation target events, (2) \textit{Are you looking at the presentation?}, detecting the contextual existence of the students by mapping the presence of the instructor and student, and (3) \textit{Are you following the presentation?}, detecting the cognitive existence of the students by comparing the instructor and the student's gazing energy (detail in Section~\ref{sec:engage_analyse}) at the screen. The next section discussed these three steps in detail. 

\section{Student Engagement Detection}\label{sec:engage_analyse}
\begin{wrapfigure}{l}{0.3\linewidth}
	\centering
	\includegraphics[clip,width=\linewidth,keepaspectratio]{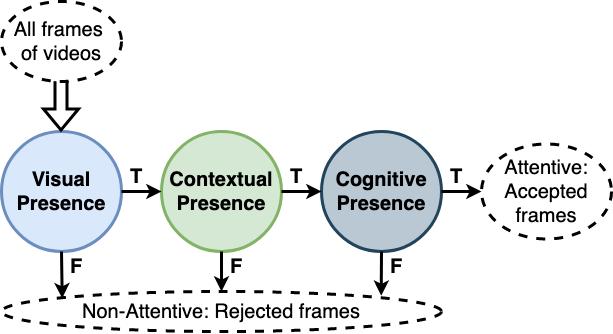}
	\caption{Schematic depiction of Student Engagement Detection mechanism flow}
	\label{fig:engageFlow}
\end{wrapfigure}
The student engagement module works on top of the fixation target extraction module to determine the students' involvement in the online class as well as the instructor's presentation performance during the online session. As we mentioned earlier, while detecting student engagement, this module responds to three questions in a cascade-line phenomenon (Figure~\ref{fig:engageFlow}), where each question module is sequentially attached. The initial module eliminates a significant number of non-engaged students based on the visual absence. Subsequent modules successively eliminate the non-engaged students with a contextual and cognitive value different from the instructor.

\subsection{Visual Presence: Are you inside the online class?}
For detecting the visual presence of the student in the online classroom, {\ourmethod} first checks whether the student's frontal face is detected during the fixation target events. {\ourmethod} detects the frontal face from the video feed\footnote{We acknowledge that capturing local video feeds may cause additional stress to the participants.} of the instructor and students using an existing approach \cite{viola2004robust} based on cascaded classifiers with Haar-like features and AdaBoost. This analysis is done one the respective devices of the students and the instructor and no data is communicated over the Internet.

\subsection{Contextual Presence: Are you looking at the lecture?}
To detect the contextual presence of the student in the class, we consider that the student visually present in the class must fixate at the presentation screen during the fixation events. This module compares the contexts of the instructor and the students in terms of the screen content. Although the student may perform different tasks during the non-fixation periods, the attentive one switches to the instructor's context during the starting of a fixation event. Therefore, for each fixation target event, the first $n$ frames from the screen capture\footnote{A screen capture records the device's (computer or laptop) screen. It can be noted that because of the privacy concern, we do not share the screen capture of one participant with another; instead, we convert it into pixel histograms which are then compared between the instructor and the students.} are chosen for making the comparison of the context. The selection of only a few initial frames from the fixation event reduces the number of comparison operations; thus, it reduces the system complexity. Considering a screen capture frame as an image, a pixel-based histogram is derived for both instructor and student-sided screens. For the student-sided presentation video, the nearby fixation target event of the instructor's presentation video is chosen. In the absence of such an event in the student-sided presentation video, the instructor's fixation target event is used. Even if both-sided presentation videos are the same, we do not observe an exact match due to the device differences. Therefore, the system scales down the histogram size to assign the nearby pixels in a single bin. Then, the system compares the histograms using the chi-square metric and selects the minimum distance among the $n$ comparisons. Finally, the student's presence in the lecture is determined depending on the distance value lying within the threshold $\delta_h$.

\begin{wrapfigure}{l}{0.3\linewidth}
	\centering
	\includegraphics[clip,width=0.6\linewidth,keepaspectratio]{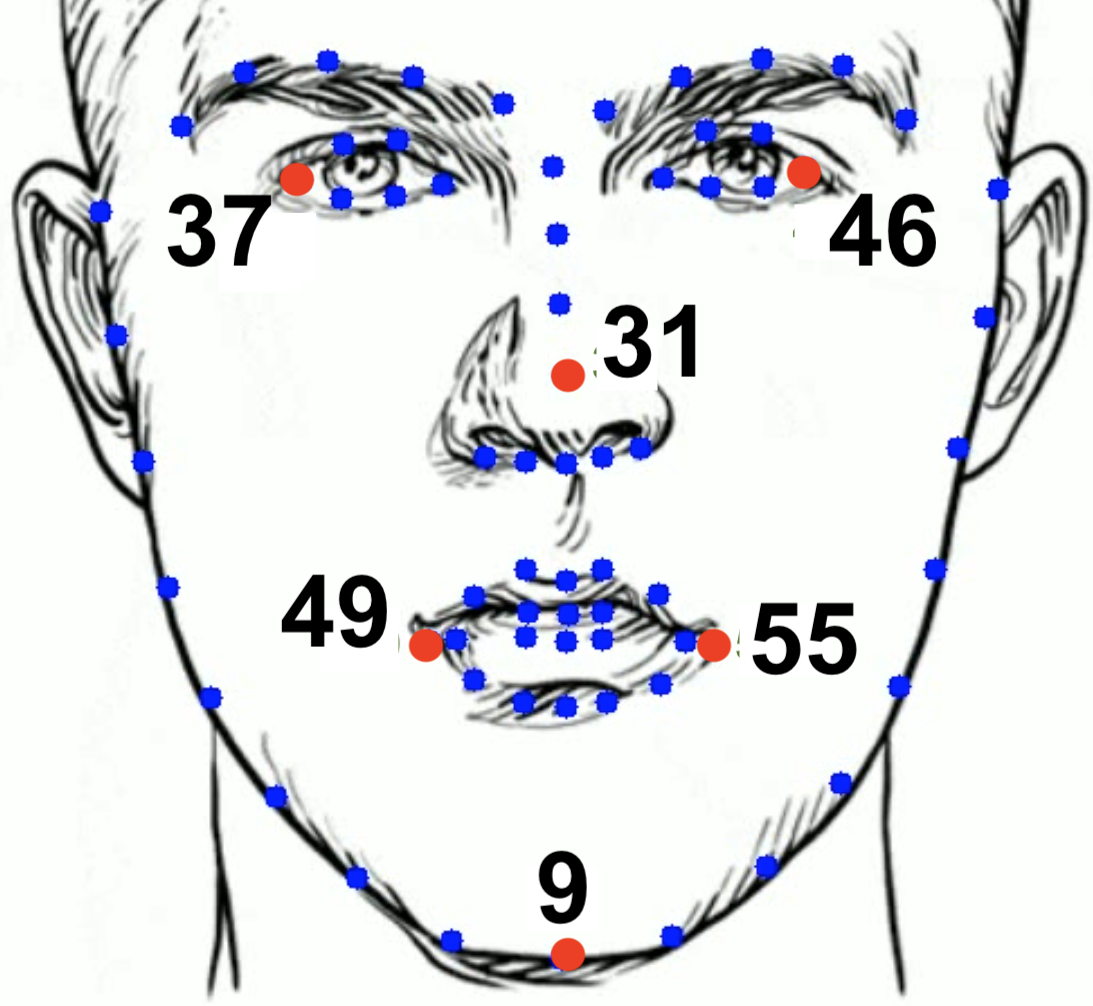}
	\caption{68 facial landmarks, candidate points for 3D points estimation (red)}
	\label{fig:facelandmark}
\end{wrapfigure}
\subsection{Cognitive Presence: Are you following the lecture?}
In this module, the system detects the cognitive presence of the students by checking whether the instructor and the student are following the presentation lecture in a similar way.  For this purpose, {\ourmethod} first detects the facial landmarks from the detected faces of the video frames following the hourglass model \cite{deng2018cascade}. From the facial region of interest, 68 facial landmarks (shown in Figure \ref{fig:facelandmark}) are generated as an outcome of the hourglass model. We next detect the gazing projection based on these facial landmarks, as follows.

\subsubsection{Gazing Projection Estimation}
This submodule estimates the position where the student gaze is projecting towards the front screen. However, due to the binocular vision problem, completely relying on the gaze for the projection is not legitimate. In the line, the eye corners move towards the direction of the eyeball. Therefore, {\ourmethod} uses facial landmarks for estimating the fixate position on the screen. In this approach, first, the 3D points in the world coordinate system for the 2D facial landmarks are determined by following an existing state-of-art mechanism \cite{wang2017real}. {\ourmethod} selects six landmarks (2 eye corners -- 37, 46; 2 lip corners -- 49, 55; 1 nose end -- 31; and 1 thin end -- 9; red points in Figure \ref{fig:facelandmark}) points out of 68 facial landmarks as the candidate points for 3D points estimation. Further, the system follows Zhang \emph{et al.} \cite{zhang2000flexible} for calibrating the camera parameters. Next, the pose of the calibrated camera is predicted from the current detected 2D landmarks, and the model populated 3D points in the world coordinate system by applying a direct linear transform solution followed by Levenberg Marquardt optimization. Without loss of generality, {\ourmethod} considers the center of the face (nose endpoint), the candidate point for the gazing projection. Finally, the system determines the projection of the nose end to the 2D screen using the current pose of the camera following the Pinhole camera model. It can be noted this this computation is done on individual devices.

\subsubsection{Gazing Energy Similarity}
From the projection point, our final task is to identify the students who are following the presentation. Towards this goal, the student's estimated projected points on the screen are compared with those of the instructor. For this purpose, the \textit{gazing energy} of the students is shared with the instructor, and the computation is done over instructor's device. While attending the lecture in the online mode, the gaze movement is highly dominated by the horizontal movement~\cite{lan2020gazegraph}. Therefore, {\ourmethod} excludes the vertical axis data for the next processing. Furthermore, as the projection value depends on various parameters like the camera calibration, and 3D-2D mapping model, the individual projection value can be erroneous. For eliminating the impact of the error, the system populates per second projection strength by computing the projection value-based gazing energy over the window of one second. The gazing energy is calculated by taking the sum of the square of the horizontal projection value over a window of one second. For engaged students, both the instructor and the student look at a similar object in the presentation. Therefore, the gazing energy must be similar for both of them. Hence, the system compares the set of gazing energy within a single fixation target event for both the instructor and the student using the Student t-test to interpret whether both the samples have a similar mean value. Our null hypothesis is that the mean of the gazing energy of the student and the instructor are the same.  The system reports non-engagement of the student depending on the $p-value < .001$.

\section{System Layout Design}
{\ourmethod} renders the online classroom involvement status in two phases. In the first phase, it computes the involvement score for both the instructor and the students. The final phase takes charge of the score generation time detection. The details of the proposed visualizer system are discussed as follows. 

\subsection{Involvement Score Computation}
The visualizer shows two types of involvement score -- (i) a current score, and (ii) an aggregate score. The current score is computed based on the individual involved in the current segment of the presentation whereas the aggregated score shows the overall involvement in the segment of the presentation. Moreover, the system displays the overall involvement in all the prior segments. 
\textbf{(1) Student Engagement Score: }
Our system preserves a positive fixation target event count $\mathcal{F}_s$ for each student $s$ to count the fixation events where the students are engaged. The fixation target event count, $\mathcal{F}_s$ is incremented by one if the cognitive presence of the student is detected in that fixation event. Therefore, for a segment of $t$ unit of time, if there exists $f$ fixation target events, then our system computes the current student engagement score as $\mathcal{C}_s = (\mathcal{F}_s/f) \times 100\%$. While counting the fixation target events, the system only considers the events where the instructor is contextually present. The aggregative score is calculated by taking the average of the current score, $\mathcal{C}_s$ of all the students present in the online class. 
\textbf{(2) Presentation Score: }
Similar to the student engagement score, {\ourmethod} computes the instructor's presentation score as a by-product of the system. Analogous to the student, for the instructor, the system maintains a positive fixation target event count $\mathcal{F}_i$ to count the fixation events where the instructor is involved. But, the fixation target event count, $\mathcal{F}_i$ is incremented by one if the contextual presence of the instructor is detected in that fixation event. Therefore, for $t$ unit time segment, if there presents $f$ fixation target events, the instructor's current presentation score is calculated as $\mathcal{C}_i = (\mathcal{F}_i/f) \times 100\%$.  

\subsection{Involvement Score Generation Time Detection}
The visualizer plays a significant role in the involvement score generation time detection. It provides two alternatives to the instructor -- (i) \textit{automatic selection}: slide-transition based time segment selection, and (ii) \textit{manual selection}: fixed time-slice based time segment selection. The details follow.   

\subsubsection{Slide Transition-based Time Segment Selection}\label{subsec:slideTran}
For automatically selecting the involvement score generation interval, our system depends on the slide transition in the presentation video. This selection process not only detects the slide transition but also eliminate the insignificant slide contents such as starting slide, ending slide, and title slide. Typically, the slide numbers are present in all the presentation slides except for the insignificant ones. Therefore, the system first locates the slide number position in the slide from the usual slide number positions --  upper right corner, lower right corner, and middle bottom of the slide. During the slide transition, the pixel values of either of the three portions reasonably change and the rest two remain the same. 
For detecting the slide transition, the system applies a $30 \times 50$ pixels grid on the three pre-defined positions of each frame of the video for cropping the portion containing the slide number. The starting slide commonly with no slide number is treated as the initial template for matching the subsequent slides. Once retrieved the cropped frame portions, the system converts that into a gray-scale image and compares the cropped image with the respective cropped image of the subsequent frame using the mean squared error metric. During the slide transition, only the comparison of the cropped images with slide numbers produces a high difference value whereas the rest of the portion comparison in transition or non-transition comparison generates almost zero difference value. Hence, by applying a simple threshold value, $\delta$ the system slices the presentation video. Once a slide transition is detected, the system further compares the frame with the initial template with no slide number using the mean square error metric. If the error value is close to zero, the system marks it as an insignificant slide and eliminates the slide portions for further processing. Otherwise, the video segment belonging to a significant slide is chosen for a segment of involvement score calculation.

\subsubsection{Time Slice-based Time Segment Selection}
The presence of the slide number in the presentation is not mandatory for an academic presentation. This leads us to open up a manual solution for selecting the involvement score generation interval. In the manual process, the system slices the presentation video based on a fixed time interval (3, 5, or 15 minutes) and computes the involvement score for that time segment.

\section{Lab-Scale Evaluation}
For understanding the effectiveness of {\ourmethod}, we first conducted a lab-scale study. The detail follows.  

\subsection{Evaluation Methodology}
Analogous to a typical classroom, the participants of the experiments are selected from a similar academic background. 
Each time, one participant performs the instructor's role, whereas the rest play the students' role. 
The instructors voluntarily choose the presentation topics. We instruct them to present content with animation, image, and highlighted text. They are open to using any presentation template. The experiments are conducted using the Google Meet platform where, both the instructor and the students use a dedicated desktop computer with a Logitech webcam c270 mounted on top of the monitor. The participants can sit at 40-60cm from the monitor under normal lighting conditions.

\subsubsection{User Details}
$13$ different participants volunteered in the lab-scaled experiment for a duration of $~15$ minutes each. During the experiment, the students are instructed to perform four different attentive and non-attentive behaviors -- (a) completely following the presentation, (b) reading an article on a different tab, (c) watching a video on a different tab, and (d) looking at the mobile. Except for the presentation, there was no restriction on the article or video content selection.  Before conducting the experiments, a self-reported communication competence form is shared among the participants. The form computes the self-reported communication competence score as per the Self-Perceived Communication Competence Scale (SPCCS) \cite{mccroskey1988self} for understanding the self-reported competence over a variety of communication contexts. The instructors are chosen depending on either completely confident or fairly confident one.

\subsubsection{Baselines}
For estimating the efficacy of {\ourmethod}, we compare it using metric-based system performance analysis under lab-scale experiments. Bace \emph{et al.} \cite{bace2020quantification} developed a visual attention detection mechanism for the mobile interaction. In this approach, the attention or the engagement is detected depending on whether the user is continuously looking at the front screen. For marking a participant as attentive, we check whether the user is looking at the screen for more than $50\%$ of the time of the class.

\subsubsection{Ground Truth Generation}\label{subsec:groundtruth}
Engagement is a subjective measure. Therefore, generating the ground truth information for evaluating the system is a difficult task. For ground truth annotation, we have asked both the instructor and the students to capture the facial and the presentation videos using the OBS\footnote{https://obsproject.com/} platform. We mark the participant as \textit{engaged} if the presentation video is opened and the participant is looking at the screen. Otherwise, the participant is marked as \textit{non-engaged}. We continue the annotation for each of the time segments of the videos. In case of a mixed behavior, we mark the participant depending on the majority behavior during that time segment.

\subsubsection{Evaluation Mechanism}\label{subsec:evalMechanism}
We select F$_{\beta}$-score for computing the efficacy of the system with unbalanced set of \textit{engaged} and \textit{non-engaged} pair. Furthermore, detecting a \textit{non-engaged} student is important for student's understandability. Therefore, we have calculated specificity and negative predictive value for prioritizing the \textit{non-engaged} detection. Specificity indicates the detected non-engaged participants by the system out of all non-engaged participants in the collected sample. In contrast, a negative predictive value indicates the detected non-engaged participants out of all detected non-engaged participants. Finally, we compute the F$_{\beta}$-score as the weighted harmonic mean of the specificity and the negative predictive value, where $\beta=2$. 
Thus, $F_{\beta} = (1+\beta^2) \frac{\makebox{negative predictive value} \times \makebox{specificity}}{\beta^2 \times \makebox{negative predictive value} + \makebox{specificity}}$.

\subsection{Results and Evaluation}
The lab-scale study gives an overall analysis of {\ourmethod} in comparison with the state-of-the-art. It further provides insight into the participants' specific performance.

\subsubsection{Baseline Comparison:}
\begin{figure}[ht]
	\begin{subfigure}{0.24\linewidth}
		\centering
		\includegraphics[clip,width=\linewidth,keepaspectratio]{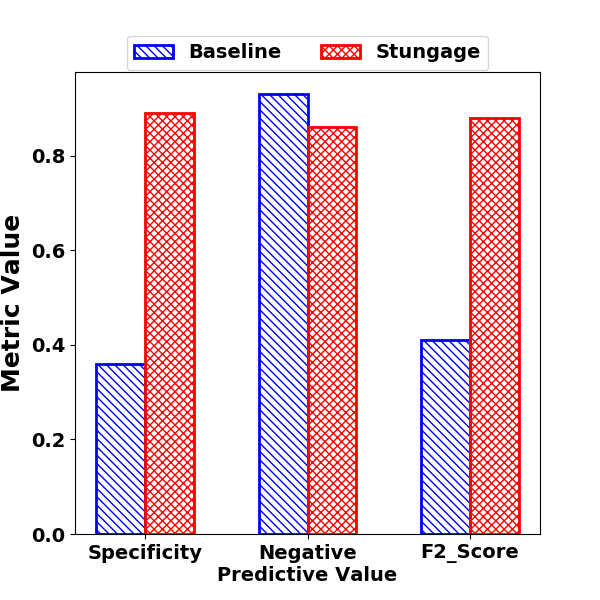}
		\caption{}
		\label{fig:baselineCompareControlled}
	\end{subfigure} 
	\begin{subfigure}{0.24\linewidth}
		\centering
		\includegraphics[clip,width=\linewidth,keepaspectratio]{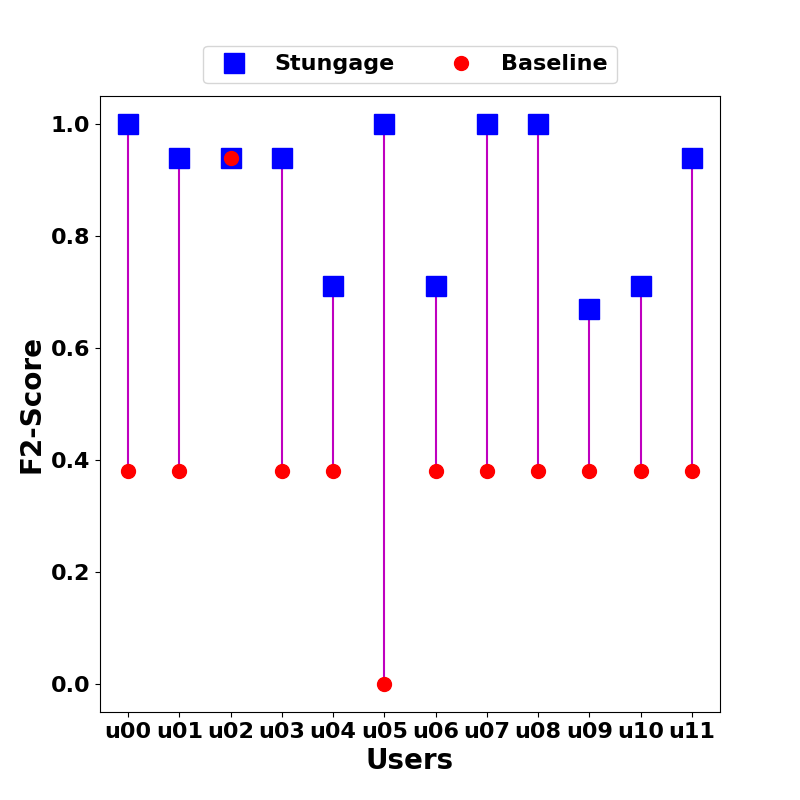}
		\caption{}
		\label{fig:userWisePerf}
	\end{subfigure} 
	\begin{subfigure}{0.24\linewidth}
		\centering
		\includegraphics[clip,width=\linewidth,keepaspectratio]{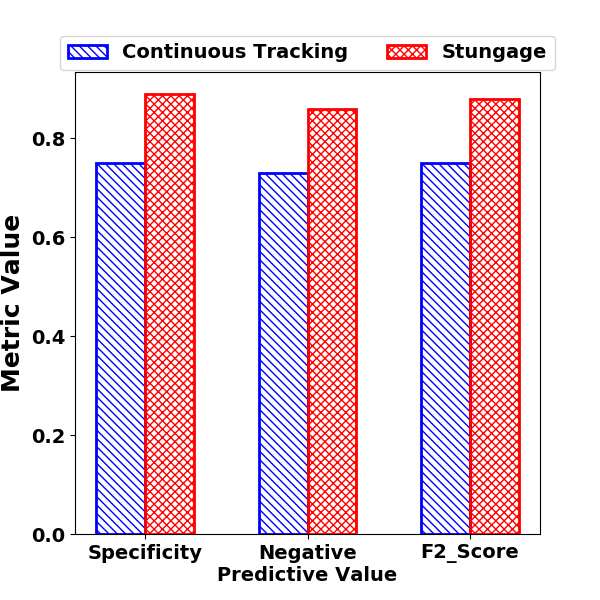}
		\caption{}
		\label{fig:ablationStudy}
	\end{subfigure} 
	\begin{subfigure}{0.24\linewidth}
		\centering
		\includegraphics[clip,width=\linewidth,keepaspectratio]{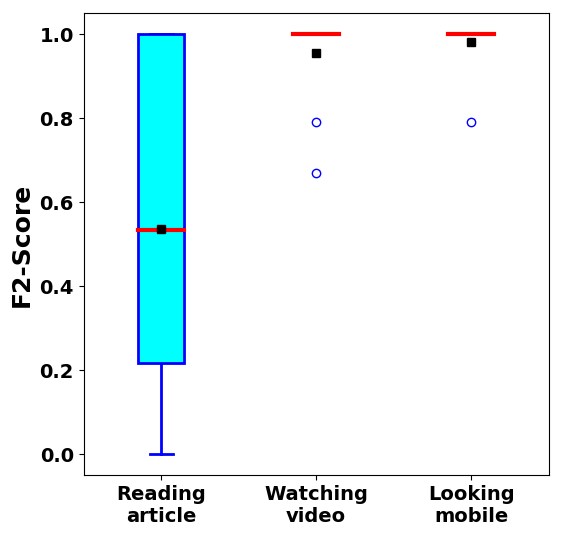}
		\caption{}
		\label{fig:attentionTypeWisePerf}
	\end{subfigure} 
	\caption{(a) System performance in lab-scale, (b) Participant wise system performance (pink line represents the difference in F2-score between {\ourmethod} and baseline), (c) Impact of the Fixation Target Extraction, (d) Different task-wise system performance}
\end{figure}

{\ourmethod} detects student engagement depending on the statistically significant probability value. For analyzing the system's efficacy in detail, we compare {\ourmethod} with the continuous monitoring-based scheme described in \cite{bace2020quantification}. Figure \ref{fig:baselineCompareControlled} shows a comparison between {\ourmethod} and the baseline. From a teacher's perspective, identifying non-attentive students is more relevant. Therefore, we use three metrics -- \textit{specificity}, \textit{negative predictive value}, and \textit{F2-score} for assessing the system performances. We observe that the \textit{F2-score} of {\ourmethod} is better than the baseline scheme under the lab-scale condition (Figure \ref{fig:baselineCompareControlled}). Although the negative predictive value is closer for both the methods, the specificity is much improved for {\ourmethod}. Even though the baseline captures the non-attentive cases (resulting in high negative predictive value), it also results in high false-positive detection (low specificity). The high false-positive cases mainly occur when the participant performs other activities on a different tab, keeping the face in front of the screen.  

\subsubsection{User-wise Performance}
For analyzing the system performance at the participants' level, we study the individual's performance in four different attentive and non-attentive behaviors. Figure \ref{fig:userWisePerf} shows the F2-score for individual participants while performing all four different behaviors using {\ourmethod} and baseline method. The figure illustrates that except for the participants $u04$, $u06$, $u09$, and $u10$, the individual F2-score of {\ourmethod} is at least $0.9$, whereas that of baseline method is $0.38$. Although the non-attentive behaviors like video watching and mobile searching are captured accurately for the participants $u04$, $u06$, $u09$, and $u10$, our system gets confused when a student reads some article on the computer screen. Indeed, such behavior is expected as the system does not explicitly differentiate between reading articles relevant to the class versus reading irrelevant articles (like a newspaper) on the computer screen.  

\subsection{Ablation Study}
We perform an ablation study where we continuously compute the students' engagement by suppressing the fixation target extraction module. Figure \ref{fig:ablationStudy} shows the system performance under both the schemes -- {\ourmethod} and continuous tracking without fixation target extraction. Irrespective of the metric value measure, the continuous tracking scheme fails to reach the performance of the complete model. The failure occurs mainly during the attentive instances when the participant takes note while attending the virtual class. Therefore, this ablation study confirms the importance of the fixation target extraction module.

\subsection{Impact of Different Tasks and Design Setup}
The types of co-tasks during multitasking play a significant role in the engagement computation as the characteristics of the student's presence in an online class highly depend on the co-task. Figure \ref{fig:attentionTypeWisePerf} shows the impact of different performing co-tasks -- (a) reading articles, (b) watching a video, and (c) looking at a mobile, during the class, on the system performance. Except for the first one, the rests are pretty different from attending a lecture. Therefore, the last two tasks get majorly excluded using the contextual and visual presence module, respectively, resulting in a high F2-score of the system. On the other side, the detection of the student's engagement while reading an article is merely symmetrical with attending the class, as both involve looking at a particular location of the screen for a significant duration. Even though the gazing projection-based cognitive computation for excluding the first task causes to generate the false positive instances, we obtain the median F2-score of $0.54$. 

\begin{figure}[ht]
	\begin{subfigure}{0.24\linewidth}
		\centering
		\includegraphics[clip,width=\linewidth,keepaspectratio]{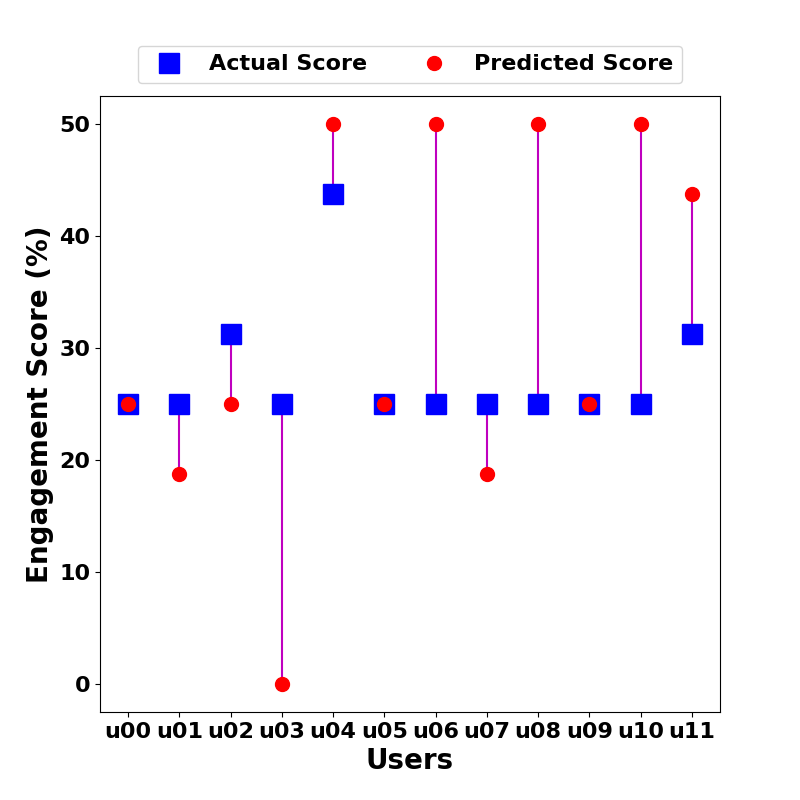}
		\caption{}
		\label{fig:userWiseScore}
	\end{subfigure} 
	\begin{subfigure}{0.24\linewidth}
		\centering
		\includegraphics[clip,width=\linewidth,keepaspectratio]{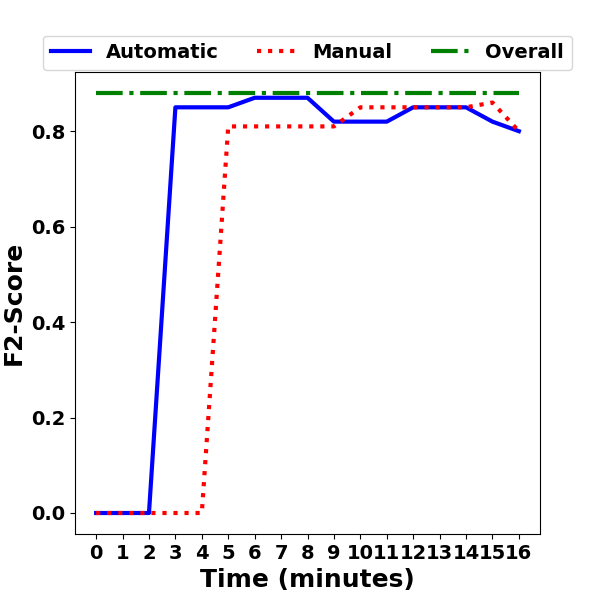}
		\caption{}
		\label{fig:timeWisePerf}
	\end{subfigure} 	
	\begin{subfigure}{0.24\linewidth}
		\centering
		\includegraphics[clip,width=\linewidth,keepaspectratio]{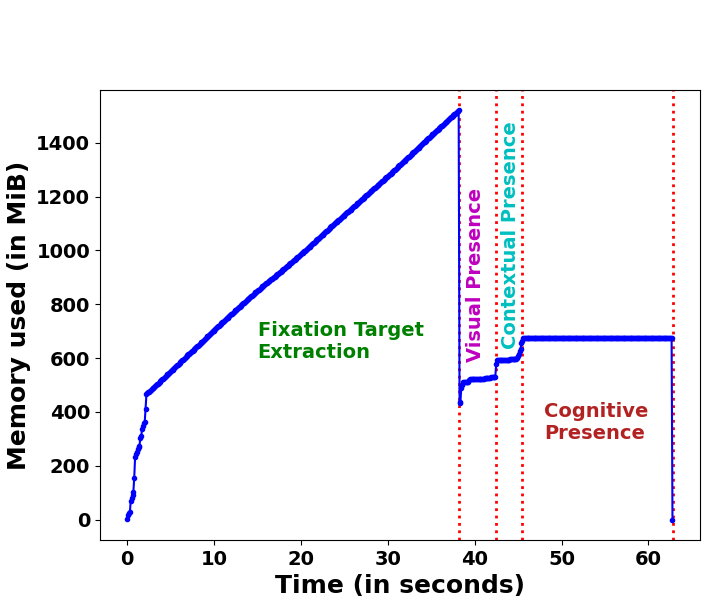}
		\caption{}
		\label{fig:memoryProf}
	\end{subfigure} 
	\begin{subfigure}{0.24\linewidth}
		\centering
		\includegraphics[clip,width=\linewidth,keepaspectratio]{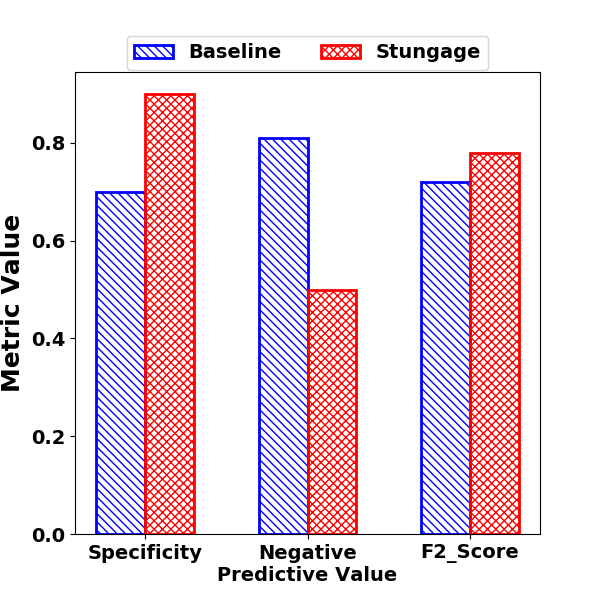}
		\caption{}
		\label{fig:baselineCompareFree}
	\end{subfigure}
	\caption{System performance: (a) participant wise (pink line represents the difference in engagement score between {\ourmethod} and baseline), (b) lecture time wise, (c) computational cost, (d) in-the-wild}
\end{figure}
Besides analyzing the module-centric impact, we further study the system performance from the design layout perspective in terms of score computation and score generation time. We observe that although the predicted score varies marginally across the participants (Figure \ref{fig:userWiseScore}), the predicted score of $67\%$ participants differs from the actual at most by $12.5(\%)$, whereas the exact match in terms of the score is found for $25\%$ of the participants. The rest of $33\%$ participants get a score to differ by $25(\%)$ due to the false positive instances caused for the article reading. For analyzing the system behavior with a varying score generation time, both automatic (slide-transition based time segment selection) and manual (5 minutes time-slice based time segment selection) time selection schemes perform almost equally (Figure \ref{fig:timeWisePerf}). Moreover, as the engagement is detected based on the statistically significant test, the overall engagement score is not the aggregation of the individual instances. However, we find that the system performance of the individual instances for both automatic (blue line) and manual (red line) are close to the overall one (green line). 
Similar to the student's engagement score, {\ourmethod} generates the instructor's presentation score with a detection error margin of $2(\%)$.

\subsection{Running Time}
For analyzing the computational cost during the system execution, we have arranged a short lecture session of $116$ seconds duration with a single instructor and student. The presentation contains three fixation target events. Figure \ref{fig:memoryProf} shows the memory consumption with time for different modules of the system. The overall system takes $63$ seconds with a maximum memory usage of $1520$MB to calculate the student engagement. Specifically, the \textit{fixation target extraction} and the \textit{cognitive presence} modules execute in $38$ and $18$ seconds with a maximum memory usage of $1520$ and $671$ MB, respectively. The computational cost is justifiable as the \textit{fixation target extraction} module executes the complete set of frames to find out the fixation events. In this arrangement, on average {\ourmethod} takes $0.018$ second to process each video frame with a per-second frame rate of $30$.

\section{In-the-wild Evaluation}
This section analyzes {\ourmethod} over an in-the-wild setup. 
Like the lab-scale setup, we performed in-the-wild experiments where the participants could use their personal devices to join the virtual classroom. We obtained the institutes' ethical committee approval for involving the students (voluntarily) in the data collection procedure for these experiments. The data have been collected over regular online classes in a university set up during the pandemic period, where the students and the instructors volunteered in the data collection procedure. We do not impose any restrictions on the sitting pattern, lighting condition, etc., to the participants during the class. $7$ instructors are chosen from $23$ participants\footnote{The participants have similar face color. Analysing the system with different face colored is a good future direction of the work.} depending on either completely confident or reasonably confident one following SPCCS~\cite{mccroskey1988self}. The experiments are conducted under $12$ different virtual classrooms with a total of $13$ hours (minimum duration: ~ $30$ minutes, maximum duration: ~ $2$ hours). On average, $8$ students from the classes have participated in these experiments. We have compared {\ourmethod} using the survey-based system design analysis to estimate the design efficacy. Three different layouts are designed based on the existing online classroom platform.  
\noindent\textit{(i) Students' static image view:} This is the default layout, where the instructor can see the lecture video along with the static images of the limited set of students. This layout allows us to compare our system in \textit{no video} and \textit{no feedback} scenarios. 
\noindent\textit{(ii) Limited set student' video view:} This is another default layout in online meeting platforms, where the instructor can see the lecture video along with a few randomly chosen students' video feeds. Note that the complete student view is not present. This layout permits us to compare our system in a limited set of student videos. 
\noindent\textit{(iii) All students' engagement view ({\ourmethod}):} In this layout, the instructor sees the lecture video and all the students' engagement statistics. The engagement stats view is initially empty and shown after the fixation target encounter. 
The ground truth is generated in a similar way as that of the controlled setup (Section \ref{subsec:groundtruth}). 
Besides the metric-based analysis, these experiments were evaluated using a set of surveys consisting of system evaluation and student understandability. Once the class is over, three different layouts are shared with the instructor, and for each layout, they were asked to fill up (1) \textbf{system evaluation survey} \cite{murali2021affectivespotlight,murali2018speaker} that captures the instructor's assessment on the system, and (2) \textbf{student understandability survey} \cite{murali2021affectivespotlight,parmar2020making} that captures instructor's experience with the view. We perform Paired Wilcoxon signed-rank tests with correction on all the survey questions to understand the differences in the survey responses across the different layouts.

Here, besides the metric-based evaluation, we focus on the system behavior study under the in-the-wild setup condition and analyze the impact of student understandability. Figure \ref{fig:baselineCompareFree} shows that similar to the controlled setup study, the F2-score of our system is better than the baseline approach. While the baseline method detects the non-attentive cases (resulting in high negative predictive value), it also has high false-positive detection (low specificity). On the other side, Table \ref{tab:stdUnderstandSurvey} shows the average responses for the questions focuses on the evaluation of the platform and the understandability of the student by the instructor as well as well-accustomed participants\footnote{Including $7$ instructors, altogether $13$ participants participated in these surveys.}, respectively. Except for the students'  privacy, {\ourmethod} layout is rated significantly higher than both the state-of-the-art systems (\textit{Lecture with limited set of student's view} and \textit{Lecture with students' static image view}) while studying the platform evaluation. A detailed analysis using Paired Wilcoxon signed-rank test reveals that in terms of students' privacy, our system is rated higher than \textit{Lecture with a limited set of student's view} ($w=5.5,p=.009$). Although no significant differences are observed in terms of system help, satisfaction, and future usability during system evaluation study, our system is rated higher than the other two layouts (\textit{Limited set of student's view} and \textit{Students' static image view}) in terms of student understandability ($w=3.0,p=.005;w=0.0,p=.002$) and presentation performance awareness ($w=3.0,p=.004;w=0.0,p=.001$). In terms of personal connection and students' response chances, {\ourmethod} is also rated higher than the \textit{Lecture with students' static image view} ($w=0.0,p=.009;w=0.0,p=.004$), respectively.

\begin{table}[]
	\caption{Mean and standard deviation for in-the-wild study survey (numbers in the brackets denote standard deviation)}
	\scriptsize
	\begin{tabular}{c|r|c|c|c}
		\hline
		\textbf{Survey}                                                                               & \textbf{Question with endpoints: "Not at all" (1) and "Very Much" (7)} & \multicolumn{1}{l|}{\textbf{\begin{tabular}[c]{@{}l@{}}Lecture with limited\\ set of student's view\end{tabular}}} & \multicolumn{1}{l|}{\textbf{\begin{tabular}[c]{@{}l@{}}Lecture with students'\\ static image view\end{tabular}}} & \multicolumn{1}{l}{\textbf{Our Design}} \\ \hline
		\multirow{5}{*}{\textit{\begin{tabular}[c]{@{}c@{}}system\\ evaluation\end{tabular}}}         & How much do you feel that the system would help you to take the class? & 4.38(1.33)                                                                                                         & 4.15(1.96)                                                                                                       & \textbf{5.31(1.9)}                      \\
		& How distracting is the system for taking the class?                    & 4.23(1.42)                                                                                                         & 1.77(1.19)                                                                                                       & \textbf{4.62(1.55)}                     \\
		& How satisfying is the system for taking the class?                     & 4.46(1.34)                                                                                                         & 3.92(1.9)                                                                                                        & \textbf{5.23(1.72)}                     \\
		& How much would you like to take future class with the system?          & 4.69(1.2)                                                                                                          & 4.08(1.94)                                                                                                       & \textbf{5.38(2.27)}                     \\
		& How much students' privacy is maintained in the system?                & 2.62(1.39)                                                                                                         & \textbf{6.77(0.42)}                                                                                              & 5.23(1.25)                              \\ \hline
		\multirow{4}{*}{\textit{\begin{tabular}[c]{@{}c@{}}student\\ understandability\end{tabular}}} & How much of a personal connection do you feel with the student?        & \textbf{4.77(2.12)}                                                                                                & 1.85(1.03)                                                                                                       & 4.08(1.82)                              \\
		& How do you feel easy to see the student understandability?             & 4.08(1.27)                                                                                                         & 1.92(1.33)                                                                                                       & \textbf{5.54(1.87)}                     \\
		& How do you feel easy to respond the student?                           & 5.23(0.97)                                                                                                         & 3.15(1.23)                                                                                                       & \textbf{5.38(0.84)}                     \\
		& How aware are you of your presentation performance?                    & 4.92(1.21)                                                                                                         & 2.23(0.89)                                                                                                       & \textbf{6.23(0.7)}                      \\ \hline
	\end{tabular}
	\label{tab:stdUnderstandSurvey}
\end{table}
\section{Usability Study}
For the usability study, we have created a detailed demo\footnote{\url{https://youtu.be/2eUVEoKKEpU}} of {\ourmethod} containing the system running steps and then made it publicly available along with the platform. The users were free to check the system and provide their feedback through a Google form. 
The feedback form consists of $10$ questions from the System Usability Scale \cite{brooke1996sus} where the participants need to rate the system on a scale of 1 (strongly disagree) to 5 (strongly agree). 
The details of this questionnaire are available at~\cite{brooke1996sus}. Out of the $10$ questions, the odd and the even questions yield strong agreement and disagreement, respectively, for the high usability of a system. Each question's score contribution is a map to the range between 0 and 4. The overall value of system usability is calculated as,

\noindent$SU = ((Q1-1)+(5-Q2)+(Q3-1)+(5-Q4)+(Q5-1)+(5-Q6)+(Q7-1)+(5-Q8)+(Q9-1)+(5-Q10)) \times 2.5$. 

\noindent We obtain $92$ responses with the majority of the participants ($57\%$) having the age group of $25$-$35$. Besides teachers and professors, we also get responses from high school students, undergrad students, and IT professionals. The participants confirmed that they use such meeting platforms regularly for attending classroom lectures or public tutorials.

\begin{wrapfigure}{l}{0.48\linewidth}
	\begin{subfigure}{0.48\linewidth}
		\centering
		\includegraphics[clip,width=\linewidth,keepaspectratio]{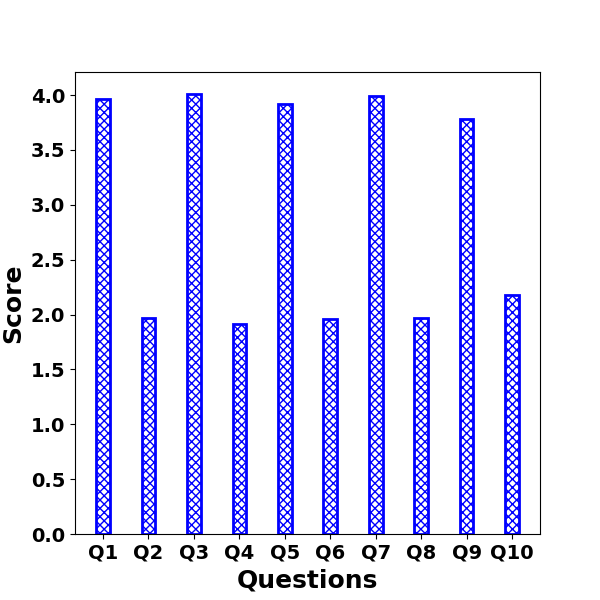}
		\caption{}
		\label{fig:quesSusScore}
	\end{subfigure} 
	\begin{subfigure}{0.48\linewidth}
		\centering
		\includegraphics[clip,width=\linewidth,keepaspectratio]{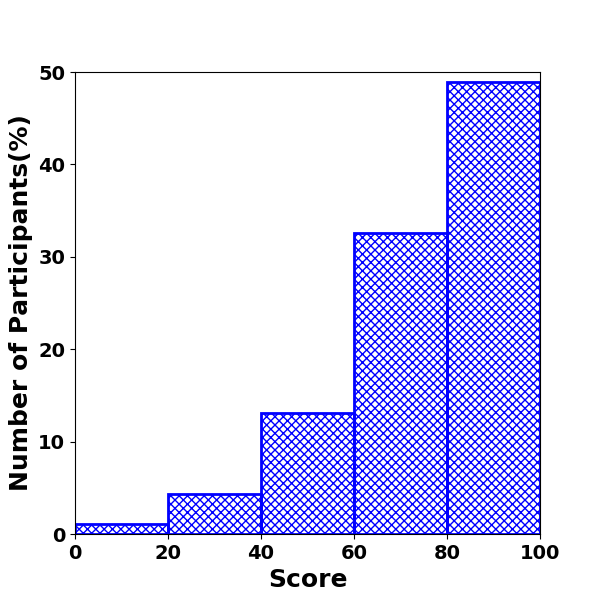}
		\caption{}
		\label{fig:participantSusScore}
	\end{subfigure} 
	\caption{Statistical analysis of SUS: (a) question-wise, (b) participant-wise}
	\label{fig:sus}
\end{wrapfigure}

For establishing the usefulness of our system, we check the SUS score distribution from the public feedback. Figure~\ref{fig:quesSusScore} shows the individual question-wise SUS score which confirms that the participants provide their feedback by properly reading the instruction, concluding that they are valid users. On the other side, Figure~\ref{fig:participantSusScore} reveals that $49\%$ of the participants have given the SUS score of more than $80$ whereas the average SUS score is $74.18$. This indicates that the participants in the survey consider {\ourmethod} as a useful system for understanding the students' engagement. 

Besides the usability questions, we keep an optional open-ended suggestion field in the feedback form. This results in receiving a few inspiring words along with appreciation from the participants. One of the participants mentioned ``\textit{It looks like an interesting application to me. But the efficiency of the facial recognition code needs to be tested properly.}'' Truly, as our system uses various existing computer vision tools for processing facial as well as presentation videos, the system performance utterly depends on the efficacy of those tools. We receive justifiable system performance under state-of-the-art tools. Further improvement of those tools will promote our system. Another valuable suggestion is -- ``\textit{Real-time interactions like pop up questions and random opinion taking may be incorporated in the student interface alongside the instructor video and content presentation.}'' Here, the system only captures the current students' involvement status during the online class without involving them. Adding a recommender for improving the current students' involvement status will be interesting future work.  

\section{Conclusion}
To the best of our knowledge, {\ourmethod} is the first of its kind that identifies the discrete fixation target events followed by the visual, contextual, and cognitive presence detection for measuring the students' engagement in the virtual classroom. While quantifying the students' engagement score, we also compute the presentation score of the instructor for self-assessment. The thorough evaluation from both lab-scale and in-the-wild analysis states that {\ourmethod} performs well for the majority of the cases with good usability feedback. However, {\ourmethod} still relies on video processing, which is always a heavy task; therefore, it would be interesting to optimize the system further to make it more suitable for handheld devices.

\bibliographystyle{ACM-Reference-Format}
\bibliography{ref/refernce}

\end{document}